\documentclass[aps,prb,twocolumn,unsortedaddress,showpacs,preprintnumbers]{revtex4}

\usepackage{graphicx}

\begin{document}

%\preprint{To be published in Physical Review B}

\title{A Universal Intrinsic Scale of Hole Concentration for High-$T_c$ Cuprates}

\author{T. Honma}
\email[E-mail address : ]{homma@asahikawa-med.ac.jp}
\affiliation{Department of Physics and Texas Center for Superconductivity and Advanced Materials, University of Houston, Houston, TX77204-5005, U.S.A.}
\affiliation{Department of Physics, Asahikawa Medical College, Asahikawa 078-8510, Japan}

%\author{P.H. Hor,$^1$ and H.H. Hsieh$^{1,2}$}
%\affiliation{$^1$Department of Physics and Texas Center for Superconductivity and Advanced Materials, University of Houston, Houston, TX77204-5005, U.S.A.}
%\affiliation{$^2$Department of Electrical Engineering, Chung Cheng Institute of Technology, National Defense University, Taoyuan, 335 Taiwan, R.O.C.}

\author{P.H. Hor}
\affiliation{Department of Physics and Texas Center for Superconductivity and Advanced Materials, University of Houston, Houston, TX77204-5005, U.S.A.}

\author{H.H. Hsieh}
\affiliation{Department of Physics and Texas Center for Superconductivity and Advanced Materials, University of Houston, Houston, TX77204-5005, U.S.A.}
\affiliation{Synchrotron Radiation Research Center, Hsinchu, Taiwan 30077, Republic of China}

\author{M. Tanimoto}
\affiliation{Department of Physics, Asahikawa Medical College, Asahikawa 078-8510, Japan}
\date{\today}

\begin{abstract}
We have measured thermoelectric power (TEP) as a function of hole concentration per CuO$_2$ layer, $P_{pl}$, in Y$_{1-x}$Ca$_x$Ba$_2$Cu$_3$O$_6$ ($P_{pl}$ = $x$/2) with no oxygen in the Cu-O chain layer. The room-temperature TEP as a function of $P_{pl}$, $S^{290}$($P_{pl}$), of Y$_{1-x}$Ca$_x$Ba$_2$Cu$_3$O$_6$ behaves identically to that of La$_{2-z}$Sr$_z$CuO$_4$ ($P_{pl}$ = $z$). We argue that $S^{290}$($P_{pl}$) represents a measure of the intrinsic equilibrium electronic states of doped holes and, therefore, can be used as a common scale for the carrier concentrations of layered cuprates. We shows that the $P_{pl}$ determined by this new universal scale is consistent with both hole concentration microscopically determined by NQR and the hole concentration macroscopically determined by the formal valency of Cu. We find two characteristic scaling temperatures, $T_S^*$ and $T_{S2}^*$, in the TEP vs. temperature curves that change systematically with doping. Based on the universal scale, we uncover a universal phase diagram in which almost all the experimentally determined pseudogap temperatures as a function of $P_{pl}$ fall on two common curves; $lowe$r $pseudogap$ temperature defined by the $T_{S}^*$ versus $ P_{pl}$ curve and $upper$ $pseudogap$ temperature defined by the $T_{S2}^*$ versus $P_{pl}$ curve. We find that while pseudogaps are intrinsic properties of doped holes of a single CuO$_2$ layer for all high-$T_c$ cuprates, $T_c$ depends on the number of layers, therefore the inter-layer coupling, in each individual system.

\end{abstract}


\maketitle

\section{INTRODUCTION}

Understanding some of the peculiar properties of the high temperature superconductors (HTSC's) is one of the challenging problems in the condensed matter physics. Especially, in the underdoped region, not only the unusually high superconducting transition temperature ($T_c$) but also many normal state properties have defied our current knowledge of metal. The tremendous experimental results have been accumulated since since the discovery of HTSC in 1986. Unfortunately, due to both the experimental and material constrains, many high quality data were collected on different materials. For instance, neutron scattering experiments require very large single crystals and therefore have almost exclusively been done on YBa$_2$Cu$_3$O$_y$ (Y123) and La$_{2-z}$Sr$_z$CuO$_4$ (LS214) that big crystals can be grown. Angle-resolved photoemission spectroscopy (ARPES) and scanning tunneling microscopy (STM), which are sensitive to surface conditions, have been performed mainly on Bi$_2$Sr$_2$CaCu$_2$O$_y$ that a virgin surface can be obtained by cleavage. On the other hand, resistivity $\rho$, Hall coefficient $R_H$ and thermoelectric power (TEP) $S$ have been measured on almost all HTSC's. It will be most fruitful if different measurements done on different materials can be compared and analyzed on a common ground, say, hole concentration per CuO$_2$ layer, $P_{pl}$. $P_{pl}$ can be determined by Sr content $z$ ($P_{pl}$ = $z$) in La$_{2-z}$Sr$_z$CuO$_4$ with only cation doping, but it is hard to determine $P_{pl}$ unambiguously in the other systems with anion doping. If $P_{pl}$ can be determined by either one of $\rho$, $R_H$ or $S$, then almost all the available data can be compared quantitatively and analyzed on a common physical ground. Especially, establishing a scale based on TEP is much powerful, since the $S$ of HTSC is not sensitive to both the grain boundary effect and the porosity effect.\cite{carrington94} Therefore, with some precautions, both single crystal and polycrystalline data can be compared.

The underdoped HTSC is characterized by a gap-like anomaly appeared below a characteristic temperature, the so-called pseudogap temperature. The pseudogap behavior was first observed in $^{89}$Y NMR Knight shift $\Delta$$K$ and in the $^{63}$Cu spin-lattice relaxation rate $^{63}$($T_1T$)$^{-1}$.\cite{yasuoka89} It also showed up as an anomaly in the $\rho$ and $R_H$ versus temperature curves.\cite{ito93,hwang94} Subsequently, the pseudogap behavior has been observed in many experimental probes, such as far infrared,\cite{basov94} ARPES,\cite{ding96} TEP,\cite{bernhard96} specific heat,\cite{loram93,loram98} time resolved quasiparticle relaxation (QPR) measurement\cite{kabanov99} and so on. The pseudogap temperature systematically varies with cation or anion doping in individual systems. However, due to the above reason, these pseudogap temperatures can not be compared based on the hole concentration. In such a situation, the room temperature TEP ($S^{290}$) was proposed to be useful for determining $P_{pl}$.\cite{obertelli92} According to this scale, the pseudogap behavior is summarized on the phase diagram that the pseudogap temperature falls from higher temperature at lower hole concentration to zero at a critical hole concentration across the $T_c$ curve.\cite{tallon01} On the other hand, the result of ARPES suggests that the pseudogap temperature does not cross the $T_c$ curve, but smoothly merges with $T_c$ on the slightly overdoped side.\cite{ding96} 

For LS214 system, $T_c$ appears at $P_{pl}$ $\approx$ 0.06, passes a maximum $T_c$ ($T_c^{max}$) at $P_{pl}$ $\approx$ 0.16, and finally falls to zero at $P_{pl}$ $\approx$ 0.27. It was approximated by a parabolic curve\cite{presland91}
\begin{eqnarray}
 T_c/T_c^{max}=
 \left.
 \begin{array}{ll}
 1-82.6(P_{pl}-0.16)^2. & \label{form1}
\end{array}
\right.
\end{eqnarray}
Although Formula (1) can be conveniently used to estimate $P_{pl}$ for the systems with similar parabolic variation of $T_c$, it cannot be used for the systems with a complex variation of $T_c$ such as Y123. $P_{pl}$ of Y123 was estimated from the bond valence sum (BVS) analysis, which relied on accurate knowledge of interatomic bond lengths.\cite{tallon90} Later, it was shown that, for some HTSC's, $S^{290}$ can be used to estimate $P_{pl}$ consistent with that determined by either Formula (1) or BVS analysis.\cite{obertelli92} The following empirical formula has been proposed.\cite{tallon95} 
\begin{eqnarray}
S^{290}[\mu V\!\!/\!K]\!=\!\!
\left\{
    \begin{array}{ll}
    \!\!372\exp(-32.4P_{pl}) & \mbox{$(0.00\!<\!P_{pl}\!<\!0.05)$} \\
    \!\!992\exp(-38.1P_{pl}) & \mbox{$(0.05\!<\!P_{pl}\!<\!0.155)$} \label{form2} \\
    \!\!24.2-139P_{pl} & \mbox{$(0.155\!<\!P_{pl})$} .
    \end{array}
\right.
\end{eqnarray}

Formula (2) has been widely used with the distinct advantages that $S^{290}$ is material independent. Therefore, it can be used to compare physical properties as a function of $P_{pl}$ of very different HTSC's. But, there are still difficulties in using Formula (2). For instance, there is no reason why the optimal $P_{pl}$, where $T_c^{max}$ appears, should be universally $\sim$ 0.16 for HTSC as has already been questioned in Ref.\ \onlinecite{markiewicz02}. It is also not trivial to apply BVS to determine $P_{pl}$ for systems with internal strain due to the CuO chain structure such as Y123.\cite{goldschmidt93} Furthermore it is reported that, for LS214,\cite{obertelli92} (Ca$_x$La$_{1-x}$)(Ba$_{1.75-x}$La$_{0.25+x}$)Cu$_3$O$_y$ (C$_x$LBLC), \cite{knizhnik99} and Bi$_2$Sr$_{2-x}$La$_x$CuO$_y$ (Bi2201),\cite{ando00} $P_{pl}$'s determined by Formula (1) are not consistent with those determined by Formula (2). 

Y$_{1-x}$Ca$_x$Ba$_2$Cu$_3$O$_6$ (YC1236) without CuO chain has two equivalent CuO$_2$ planes, therefore the $P_{pl}$ can be determined unambiguously by Ca content $x$ as $P_{pl}$ = $x$/2. Different from the well known and widely used empirical correlation between $S^{290}$ and $P_{pl}$,\cite{obertelli92} we find that $S^{290}$($P_{pl}$) for YC1236 behaves identically to that for LS214. Since the crystal structure for YC1236 is very different from that for LS214, we argue that this new correlation of $S^{290}$ can be used as an intrinsic scale of $P_{pl}$ for different HTSC cuprates. We demonstrate that this conjecture seems to work by uncovering a universal phase diagram of the pseudogap and superconductivity. We find that while the pseudogap phase is an intrinsic property to single CuO$_2$ layer, the bulk $T_c$ seems to be governed exclusively by the inter-layer coupling. 

\section{EXPERIMENTAL}

\begin{figure}[t]
\includegraphics[scale=1]{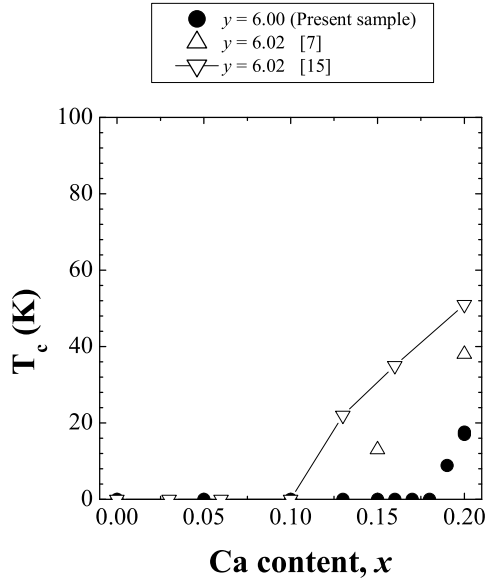}
\caption{\label{fig1}Superconducting transition temperature ($T_c$) as a function of Ca doping level for the double-layer Y$_{1-x}$Ca$_x$Ba$_2$Cu$_3$O$_y$. The data for the samples with $y$ = 6.00 are present work. The others are data reported in the literature.\cite{bernhard96,tallon95} The solid lines are guide to the eyes. }
\end{figure}

Y$_{1-x}$Ca$_x$Ba$_2$Cu$_3$O$_6$ with different Ca contents of $x$ = 0.05, 0.10, 0.13, 0.15, 0.17, 0.18, 0.19, 0.20, and 0.22 were prepared by causing a solid-state reaction in a proportioned mixture of Y$_2$O$_3$ (5N), CaCO$_3$ (5N), BaCO$_3$ (5N), and CuO (4N). These powders were ground, pressed and fired in flowing O$_2$ at 900 $^\circ$C for 6 h. This process was repeated several times. For the final firing, two pellets with $\sim$ 0.5 g each were fired for 10 $\sim$ 15 h in flowing O$_2$ at 930 $\sim$ 940 $^\circ$C. The O$_2$ gas was exchanged into Ar gas (99.9995 $\%$) at the high temperature, before in the furnace with flowing Ar gas the samples were annealed at 750 $^\circ$C and cooled to room temperature. The oxygen content $y$ was confirmed to be 6.00 $\pm$ 0.01 by using an iodometric titration technique under Ar gas. The prepared Y$_{1-x}$Ca$_x$Ba$_2$Cu$_3$O$_6$ samples with $x$ $\leq$ 0.22 were identified as a single-phase by examining the X-ray powder diffraction pattern. The Y$_{0.75}$Ca$_{0.25}$Ba$_2$Cu$_3$O$_6$ showed some minor second phase, although the main peak of the second phase was below 1 $\%$ for the main peak of the 123 phase. The density of all prepared samples was over 80 $\%$ of the theoretical density. Figure\ \ref{fig1} shows the Ca content dependence of $T_c$ with previous reported results.\cite{bernhard96,tallon95} The prepared samples show no clear superconducting transition until $x$ = 0.18, while the superconductivity appears above $\sim$ 0.125 in some other group's samples with $y$ = 6.02.\cite{bernhard96,tallon95} Accordingly, the chain-site oxygen of the present samples is confirmed to be adequately reduced. The TEP was measured by an ac method with a low-frequency (33 mHz) heating technique.\cite{choi01}

\section{RESULTS AND DISCUSSION}

\subsection{Temperature dependence of TEP for Y$_{1-x}$Ca$_x$Ba$_2$Cu$_3$O$_6$}

\begin{figure}[b]
\includegraphics[scale=1.1]{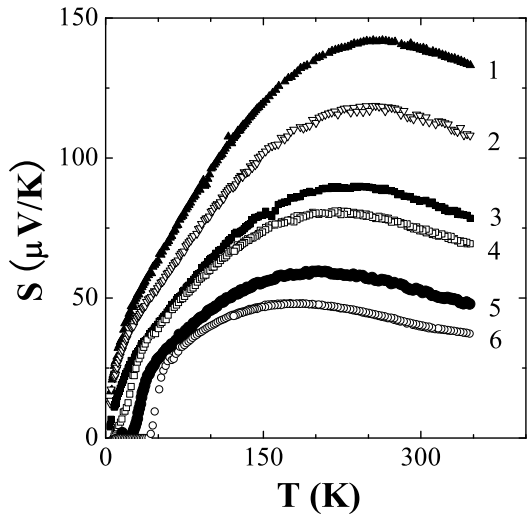}
\caption{\label{fig2} The evolution of $S$($T$) for different Ca contents of the double-layer Y$_{1-x}$Ca$_x$Ba$_2$Cu$_3$O$_6$. The Ca contents of samples 1-6 are 0.10, 0.13, 0.15, 0.18, 0.20, and 0.22, respectively. }
\end{figure}

\begin{figure}[t]
\includegraphics[scale=1.1]{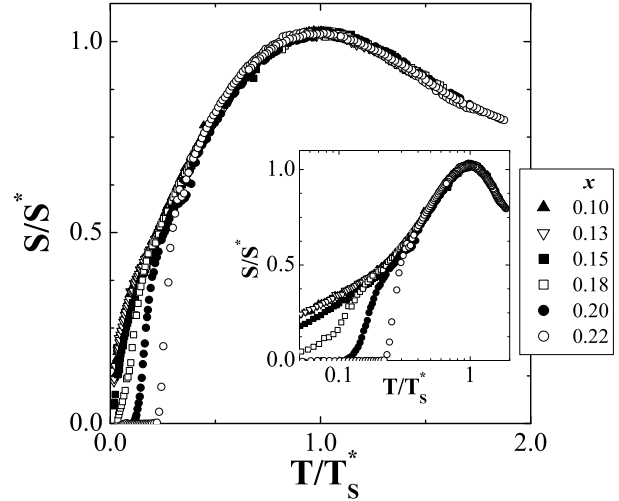}
\caption{\label{fig3} $S$/$S^*$ as a function of $T$/$T_S^*$ for different Ca contents of the double-layer Y$_{1-x}$Ca$_x$Ba$_2$Cu$_3$O$_6$. The inset shows $S$/$S^*$ vs $T$/$T_S^*$ on the logarithmic scale. }
\end{figure}

Figure\ \ref{fig2} shows the typical $T$-dependence of TEP for a series of fully deoxygenated Y$_{1-x}$Ca$_x$Ba$_2$Cu$_3$O$_6$ with 0.05 $\leq$ $x$ $\leq$ 0.22. The oxygen content is fixed to be 6.00 $\pm$ 0.01 here. Typically, upon increasing temperature $T$, positive $S$ rises towards a broad peak at a temperature $T_S^*$ and $S$ decreases almost linearly for $T$ $>$ $T_S^*$. $T_S^*$ was reported to be closely related with pseudogap temperature.\cite{bernhard96} The $T$-dependence of TEP, $S$($T$), systematically changes with Ca doping. The $T_S^*$ becomes lower with Ca doping and the magnitude of TEP decreases. The observed $S$($T$) is very similar to that reported in LS214,\cite{nakamura93,nishikawa94,zhou95,kakinuma99,johnston87,xu00,hongjie01} C$_{0.4}$LBLC,\cite{knizhnik99} and CaLaBaCu$_3$O$_y$ (CLBC)\cite{hayashi96} without ordered CuO chain. Thus, it is confirmed that there is no significant chain contribution to TEP.\cite{bernhard96} 

$S$($T$) can be well scaled by the value $S^*$ and temperature $T_S^*$ of the peak as shown in Fig.\ \ref{fig3}. $S$/$S^*$ can be fitted to a log$T$ law for 0.3 $<$ $T$/$T_S^*$ $<$ 0.8 as shown in the inset. We confirmed that $S$($T$) data of Bi$_2$Sr$_2$Ca$_{1-x}$Y$_x$Cu$_2$O$_y$ (Bi2212),\cite{akoshima98} CLBC,\cite{hayashi96} and C$_{0.4}$LBLC \cite{knizhnik99} also show the same scaling behavior by using the reported $S$($T$) data. Log$T$ behavior for $T$ $<$ $T_S^*$ was reported in CLBC\cite{hayashi96} and YBa$_2$Cu$_3$O$_y$ ($y$ $\le$ 6.48).\cite{obertelli92} Similar scaling behavior was also reported in the single-layer LS214, double-layer Y123 ($y$ $\le$ 6.65) and Bi2212.\cite{cooper96,mandal96} Since all the above observations were done on samples without ordered CuO chains like Y123, we conclude that the broad peak at $T_S^*$, log$T$ dependence for $T$ $<$ $T_S^*$ and $T$ linear behavior for $T$ $>$ $T_S^*$ are intrinsic characteristic TEP properties of doped holes in the CuO$_2$ layer. As the superconductivity appears with doping, the scaled curve of $S$/$S^*$ versus $T$/$T_S^*$ can be cut off at low temperature side as shown in Fig.\ \ref{fig3}. Accordingly, in the highly doped samples, the scaled curve for TEP could loss the log$T$ behavior by the developement of the superconductivity. The schematic picture was shown in Fig. 1 of Ref.\ \onlinecite{bernhard96}. In fact, such variation of $S$($T$) with doping was observed in Bi2212.\cite{mandal96} 

In HgBa$_2$CuO$_{4+\delta}$ (Hg1201), the $S$($T$) was well scaled by the temperature, where starts to decrease linearly with increasing $T$, and the value at the temperature.\cite{yamamoto00} The similar scaling was reported also in Bi2212.\cite{takemura00} Another characteristic temperature $T_{S2}^*$ ($>$ $T_S^*$) for TEP is reported in Zn-substituted Y123.\cite{tallon95b} The $S$($T$ $<$ $T_{S2}^*$) is suppressed by Zn substitution, while the $S$($T$ $>$ $T_{S2}^*$) does not depend on Zn substitution. Recently, the temperature used for another scaling method is reported to be just equal to the $T_{S2}^*$.\cite{yamamoto02} Therefore, the $S$($T$) can be characterized by two temperatures, $T_S^*$ where $S$($T$) has a maximum and $T_{S2}^*$ where $S$($T$) becomes sensitive to Zn substitution. We will discuss that both $T_S^*$ and $T_{S2}^*$ are related to the pseudogap in the following section.

\subsection{Room temperature (RT) scale for hole concentration}

\begin{figure}[t]
\includegraphics[scale=1.4]{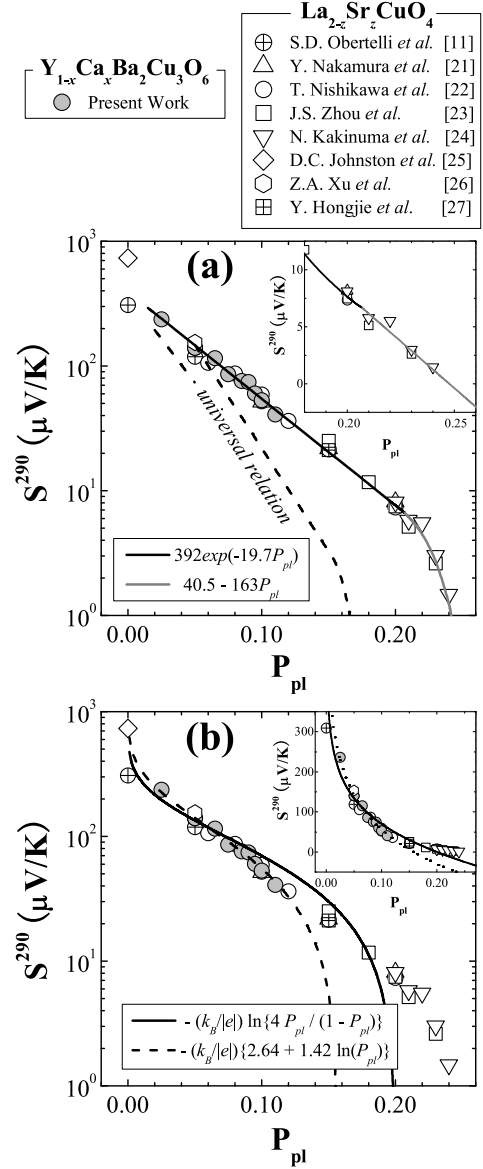}
\caption{\label{fig4} $S^{290}$ on the logarithmic scale versus $P_{pl}$ for YC1236 and LS214. Inset: $S^{290}$ on the linear scale vs $P_{pl}$. (a) The long dashed line shows the Formula (2).\cite{obertelli92} The black solid line and gray solid line show the Formula (3a) and (3b), respectively.  (b) The solid line shows the Formula (4).\cite{cooper87} The dashed line shows the Formula (5b) at 290 K.\cite{nagaosa90} }
\end{figure}

Figure\ \ref{fig4}(a) shows $S^{290}$ on the logarithmic scale as a function of $P_{pl}$. The closed circles represent the $S^{290}$ for the fully deoxygenated samples of Y$_{1-x}$Ca$_x$Ba$_2$Cu$_3$O$_6$ with 0.05 $\leq$ $x$ $\leq$ 0.22. The $P_{pl}$ can be determined unambiguously by Ca content $x$ as $P_{pl}$ = $x$/2, since YC1236 has two equivalent CuO$_2$ planes without CuO chain. For the comparison, the reported results of $S^{290}$ in LS214 are represented by various different symbols\cite{obertelli92,nakamura93,nishikawa94,zhou95,kakinuma99,johnston87,xu00,hongjie01} and the universal curve proposed in Ref.\ \onlinecite{tallon95} is shown as the long dashed line. We find that the observed log($S^{290}$) of YC1236 varies linearly with $P_{pl}$ and does not follow the universal line. This new relation can be represented by Formula (3a). It can be clearly seen that $S^{290}$ of La$_{2-z}$Sr$_z$CuO$_4$ with 0.05 $\leq$ $z$ $\leq$ 0.21 fall exactly on the same curve as $S^{290}$($P_{pl}$) for YC1236. For $P_{pl}$ $>$ 0.21, $S^{290}$($P_{pl}$) of LS214 changes from exponential to linear in $P_{pl}$ as shown in the inset of Fig.\ \ref{fig4}(a). In the pure La$_2$CuO$_4$, there is the large scattering of $S^{290}$ as shown in Fig.\ \ref{fig4}(a). $S^{290}$($P_{pl}$) of the double-layer YC1236 is identical to that of the single-layer LS214, in spite of the quite difference between YC1236 and LS214 in the crystal structure, leads us to conjecture that the present relation for $S^{290}$($P_{pl}$) can serve as a universal scale of $P_{pl}$ for layered HTSC with equivalent CuO$_2$ layers.\cite{here} 
\begin{eqnarray}
\nonumber
S^{290}[\mu V\!\!/\!K]\!=\!\!
\left\{
  \begin{array}{llr}
  \!\!392\exp(-19.7P_{pl}) & \mbox{$(0.02\!\leq \!P_{pl}\!\leq \!0.21)$} & \!(3\textrm{a}) \\
  \!\!40.5-163P_{pl} & \mbox{$(0.21\!<\!P_{pl})$} . & \!(3\textrm{b})
  \end{array}
\right.
\end{eqnarray}

In Ref.\ \onlinecite{obertelli92}, the $S^{290}$($P_{pl}$) of La$_{2-z}$Sr$_z$CuO$_4$ did not follow Formula (2). Two possibilities were pointed out.\cite{obertelli92} The deviation may arise from scattering effects associated with the increasing concentration of oxygen vacancies within the CuO$_2$ layers which occur especially for $z$ $>$ 0.12. Structural instabilities that are related to the orthorhombic-tetragonal transition may have some effect. Since we know that YC1236 shows the tetragonal symmetry at room temperature\cite{honma97} and the oxygen content is fixed at 6.00 $\pm$ 0.01, the identical $S^{290}$($P_{pl}$) behaviors for both YC1236 and LS214 have effectively ruled out both possibilities. 

Here, we compare the $S^{290}(P_{pl})$ that we used to construct our universal scale with some theoretical works in HTSC. Firstly, J.R. Cooper et al. found that the room temperature TEP observed in LS214 can be semiquantitaively explained by a modified Heikes formula based on hopping in a strongly Coulomb correlated system\cite{cooper87}  
\begin{eqnarray}
\nonumber
 \begin{array}{lr}
 \qquad \displaystyle S(P_{pl}) = -\frac{k_B}{|e|}\ln \frac{2P_{pl}}{1-P_{pl}} -\frac{k_B}{|e|}\ln2       ,                & \qquad \quad \qquad(4)
 \end{array}
\end{eqnarray}
where $e$ is the electronic charge, and $k_B$ is Boltzmann's constant. The $S$ arises from the Heikes formula, $-$($k_B$/$|e|$)ln[$P_{pl}$/(1$-$$P_{pl}$)] and a spin entropy term, $-$($k_B$/$|e|$)ln2, with an orbital degeneracy term, $-$($k_B$/$|e|$)ln2. The experimental data up to $P_{pl}$ $\approx$ 0.2 can be represented by the Formula (4) as shown in Fig.\ \ref{fig4}(b). However, no magnetic field dependence of TEP was observed, in spite of the expectation of the disappearance of spin entropy term under magnetic field.\cite{yu88}

According to Nagaosa-Lee model using a gauge-field theory for a uniform resonating valance band (RVB) state,\cite{nagaosa90} the $S$ is represented by a sum of fermion contribution, which is proportional to $T$, and boson contribution, which is proportional to [1$-$ln(2$\pi$$P_{pl}$/$m$$k_B$$T$)]. The total $S$ can be represented by the Formula (5a). The expression can be simplified as shown in the Formula (5b).  

\begin{eqnarray}
\nonumber
 \begin{array}{llr}
 \displaystyle S(T,P_{pl}) & \displaystyle \sim \frac{k_B}{|e|}\Bigl(1-\frac{k_B}{E_{F}}T-\ln \frac{2\pi P_{pl}}{mk_BT} \Bigr) & \  (5a)\\
 \nonumber

 \\   
 \nonumber
 \displaystyle S(T,P_{pl}) & \displaystyle = \frac{k_B}{|e|}\Bigl(a_0 + a_1 T + a_2 \ln T + a_3 \ln P_{pl}\Bigr) , & (5b)
  \end{array}        
\end{eqnarray}
where $m$ is the effective mass, $a_0$, $a_1$, $a_2$, and $a_3$ are adjustable parameters. The Formula (5b) is qualitatively consistent with the observed $T$-dependence of TEP at a fixed hole concentration. The calculated TEP at 290 K, shown in Fig.\ \ref{fig4}(b), is consistent with the experimental data up to $P_{pl}$ $\sim$ 0.12. For $P_{pl}$ $>$ 0.12, the calculated $S$ is smaller than and increasingly deviates from $S^{290}$ with doping. 

Although the present Formula (3a) and (3b) are still empirical rules, it is found that the present Formula is not artificial by comparing with the calculated results of the both models. We expect that a completed theory of high $T_c$ should be able to account for the $S^{290}$ versus $P_{pl}$ curve over the whole doping region. Without even a working phenomenological theory, we resort to the same empirical approach and further check the validity of Formula (3a) and (3b) in sections C and D by showing that indeed, if applying our universal scale to different material systems, physically meaningful comparisons and conclusions can be achieved.

\subsection{Application of RT scale to Y$_{1-x}$Ca$_x$Ba$_2$Cu$_3$O$_y$} 

Many TEP data for Y$_{1-x}$Ca$_x$Ba$_2$Cu$_3$O$_y$ have been reported.\cite{obertelli92,bernhard96,tallon95b,cooper96,cooper91,akoshimab98,cooper00,wang01} The anisotropy of the in-plane resistivity becomes significant when $y$ $\approx$ 6.68 in YBa$_2$Cu$_3$O$_y$.\cite{ito93} Therefore we assume that the Y123 below $y$ $\approx$ 6.68 does not have the conductive chain or chain contribution to the transport property. Consequently, for the Y123 systems, the RT scale for the TEP was applied to the samples up to $y$ $\approx$ 6.68. 

\begin{figure}[b]
\includegraphics [scale=1.15] {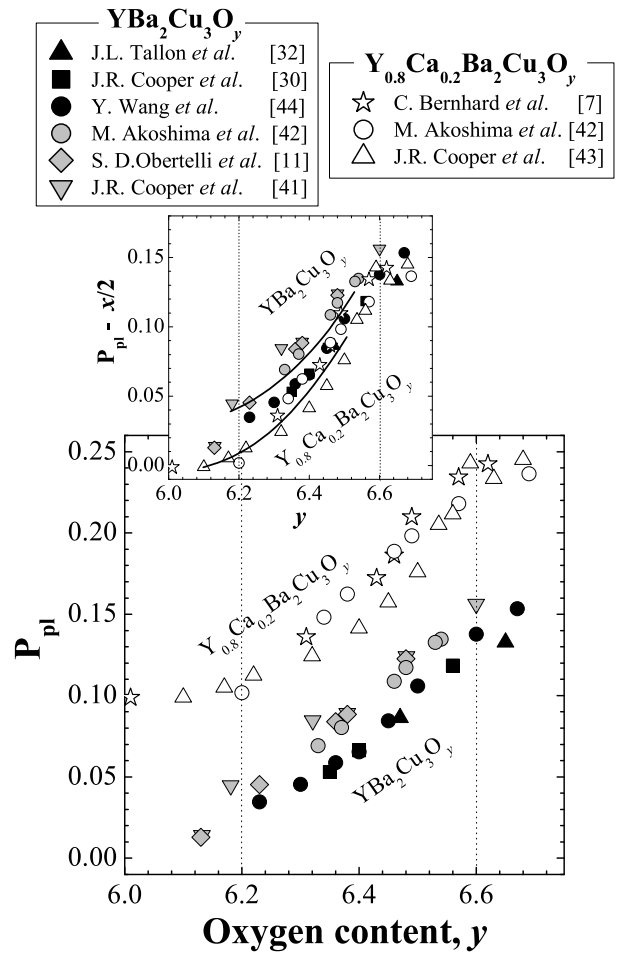}
\caption{\label{fig5} $P_{pl}$ determined from $S^{290}$ as a function of oxygen-content, $y$ in Y$_{1-x}$Ca$_x$Ba$_2$Cu$_3$O$_y$. The inset shows the effective hole concentration ($P_{pl}$$-$$x$/2) by the oxygen doping alone. The solid lines are guide to the eyes. }
\end{figure}

Figure \ref{fig5} shows the $P_{pl}$ determined from $S^{290}$ as a function of oxygen content for Y$_{1-x}$Ca$_x$Ba$_2$Cu$_3$O$_y$. The $P_{pl}$'s for YBa$_2$Cu$_3$O$_y$ are plotted by the closed black/gray symbols. The $P_{pl}$'s for Y$_{0.8}$Ca$_{0.2}$Ba$_2$Cu$_3$O$_y$ are plotted by the open symbols. In YBa$_2$Cu$_3$O$_y$, the hole carrier appears by oxygen doping beyond $y$ $\approx$ 6.15. There seems to be a threshold of the oxygen content for generating the hole carriers. Above $y$ $\approx$ 6.2, the hole carriers increase with oxygen doping. The $P_{pl}$($y$) curve trends to divide into two curves in the range of 6.2 $<$ $y$ $<$ 6.6. The splitting of $P_{pl}$($y$) curve may be related to the formation of the CuO chain structure or the inhomogeneity of oxygen distribution on the CuO chain site. The Ca-doped Y123 shows a slight different behavior compared to that for Y123. In Y$_{0.8}$Ca$_{0.2}$Ba$_2$Cu$_3$O$_y$, up to $y$ $\approx$ 6.2, the $P_{pl}$ is almost 0.10, equal to the hole concentration generated by Ca doping of 0.20. This suggests the hole carriers are exclusively generated through the Ca doping. Above $y$ $\approx$ 6.2, the hole carriers increase almost linearly with oxygen doping. This is due to the generation of the hole carriers by oxygen doping. But, there is no threshold behavior due to oxygen doping. The $P_{pl}$($y$) curve trends to divide into two curves like Y123. In YC1236, the hole carrier can be generated even by slight Ca doping as shown in Fig.\ \ref{fig4}. The creation of hole carrier by oxygen doping may need the adequate oxygen content or threshold of hole carrier. The inset of Fig.\ \ref{fig5} shows the effective hole concentration ($P_{pl}$$-$$x$/2) due to oxygen doping alone. Although there is some scattering, the generation of hole carrier by oxygen doping is found to be slightly suppressed in the Ca-doped samples. The orthorhombic-tetragonal transition occurs in the range of 6.3 $\alt$ $y$ $\alt$ 6.5.\cite{akoshimab98,veal90} The Ca doping may influence the formation of the microscopic CuO chain ordering. In the CLBC and C$_x$LBLC with significant Ca doping, there is no ordered CuO chain like Y123. \cite{bernhard96}

\begin{figure}[b]
\includegraphics[scale=1.15]{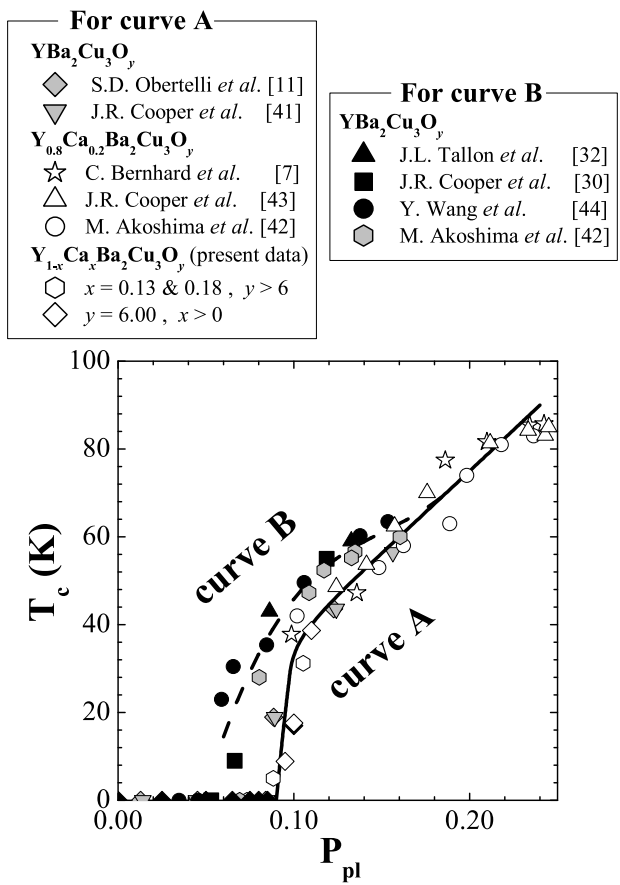}
\caption{\label{fig6} $T_c$ as a function of $P_{pl}$ in Y$_{1-x}$Ca$_x$Ba$_2$Cu$_3$O$_y$. The solid and dashed lines are guide to the eyes for the curve A and curve B, respectively. }
\end{figure}

Figure\ \ref{fig6} shows the $T_c$ as a function of $P_{pl}$ for Y$_{1-x}$Ca$_x$Ba$_2$Cu$_3$O$_y$. There are two $T_c$ vs $P_{pl}$ curves. One is that the $T_c$ appears at $P_{pl}$ $\approx$ 0.09 and increases linearly with $P_{pl}$ for $P_{pl}$ $\ge$ 0.1 (curve A). The other is that the $T_c$ appears at $P_{pl}$ $\approx$ 0.06 and merges into the curve A around $\sim$ 0.18 (curve B). The curve A includes the $T_c$($P_{pl}$) for YC1236 without CuO chain and Y$_{0.8}$Ca$_{0.2}$Ba$_2$Cu$_3$O$_y$ with $y$ $<$ 6.6. Further, the $T_c$($P_{pl}$) for CLBC and C$_x$LBLC without ordered chain also follows the curve A.\cite{knizhnik99,hayashi96} In the curve A, two data for Y123 are included. The samples for these data were prepared over 10 years ago. It is well known that tuning the oxygen content using the quenching technique often leads to a large degree of in-plane disorder which causes localization and strong $T_c$ suppression.\cite{wang01} The disorder can be removed by low-$T$ annealing, since the low-$T$ annealing causes the oxygen-rearrangement forming CuO chains.\cite{veal90} The data from the relative new Y123 samples fall on the curve B. The Y123 data on the curve A may have some disorder within the CuO chain layers, although it does not have the macroscopic chain ordering. Accordingly, the curve A is for the samples with no chain or disordered chain fragments and curve B is for the samples with the relative aligned chain fragments. In Y$_{0.8}$Ca$_{0.2}$Ba$_2$Cu$_3$O$_y$ with $y$ $\approx$ 6.5, the $P_{pl}$ is $\sim$ 0.20 and it shows the tetragonal symmetry.\cite{akoshimab98} The curve B merges into the curve A at $P_{pl}$ $\approx$ 0.18. The anisotropy of the in-plane resistivity of the pure YBa$_2$Cu$_3$O$_y$ becomes significantly above $y$ $\approx$ 6.68 ($P_{pl}$ $\approx$ 0.16),\cite{ito93} although the orthorhombic-tetragonal transition occurs at $y$ = 6.3 $\sim$ 6.5 ($P_{pl}$ = 0.05 $\sim$ 0.12).\cite{akoshimab98,veal90} Accordingly, it is considered that until $P_{pl}$ $\approx$ 0.16 the superconductivity and TEP are not influenced by the long chain ordering, at least. In spite of the same hole concentration, the $T_c$ for the curve B is slightly higher than that for the curve A. This seems to suggest that the formation of the CuO chain fragment may enhance the coupling between the CuO$_2$ planes.  

\subsection{Comparison with the hole concentration determined by other techniques: the validity of the present scale}

\begin{figure}[t]
\includegraphics[scale=1.15]{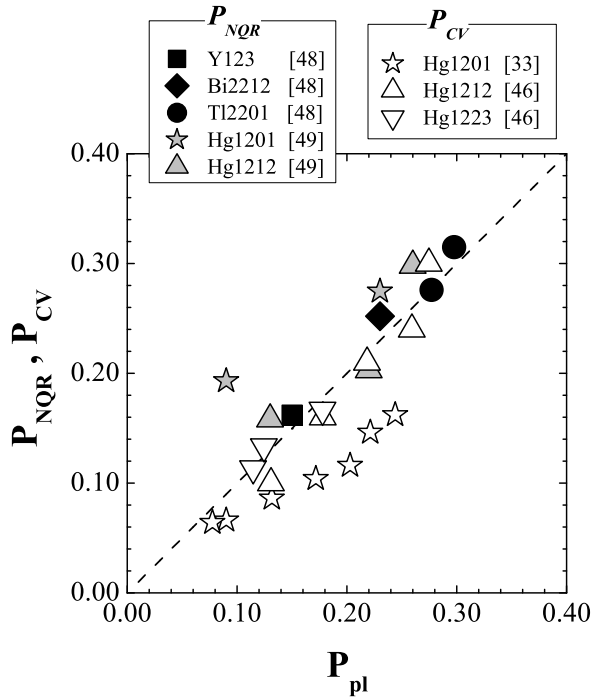}
\caption{\label{fig7} $P_{pl}$ determined from TEP by the present scale versus $P_{NQR}$ determined from NQR and $P_{CV}$ determined from Cu valency. The dashed line shows $P_{pl}$ $=$ $P_{NQR}$ $=$ $P_{CV}$. }
\end{figure}

The hole concentration per CuO$_2$ layer can be chemically estimated from the formal valancy of Cu through the titration technique. From the reported $S^{290}$ and the Cu valency for the double-layer HgBa$_2$CaCu$_2$O$_{6+\delta}$ (Hg1212) and triple-layer HgBa$_2$Ca$_2$Cu$_3$O$_{8+\delta}$ (Hg1223),\cite{yamamoto00,fukuoka97} the $P_{pl}$ and the hole concentration per CuO$_2$ layer determined from the Cu valency $P_{CV}$ are calculated. The $P_{CV}$ of Hg1212 and underdoped Hg1223 are plotted in the Fig.\ \ref{fig7} as the up- and down-ward triangle, respectively. In our scale, the optimal doping level for Hg1223 is $\sim$ 0.21 as shown in the following section. In both systems, the $P_{CV}$ is almost identical to $P_{pl}$. In the overdoped side for the triple- and four-layer HTSC, the inhomogeneity charge distribution for in-equivalent CuO$_2$ layers was reported.\cite{fujii02,tokunaga00} Since our scale was developed for the HTSC with the equivalent CuO$_2$ layer. So, we did not use our scale for the Y123 system with chain contribution to TEP. For the same reason we do not apply our scale to the overdoped triple- and four-layer systems. Noted also that, for an unknown reason, $P_{pl}$ does not coincident with $P_{CV}$ for the single-layer Hg1201. 

The hole concentration per CuO$_2$ layer for YBa$_2$Cu$_3$O$_{6.6}$, Bi$_2$Sr$_2$CaCu$_2$O$_y$ ($T_c$ = 86 K) and Tl$_2$Ba$_2$CuO$_{6+\delta}$ (Tl2201) with $T_c$ = 80 K and 42 K was estimated from the nuclear quadrupole frequency for single- and double-layer HTSC's.\cite{tokunaga99} The reported hole concentration per CuO$_2$ layer determined from the nuclear quadrupole resonance (NQR) $P_{NQR}$ is plotted in Fig.\ \ref{fig7} as a function of the $P_{pl}$. The broken line exhibits the $P_{pl}$ = $P_{NQR}$ line. The $P_{pl}$ shows a good correlation with $P_{NQR}$ in the wide doped range from 0.15 to 0.3. According to the relation between Knight shift perpendicular to the c-axis and hole concentration for single- and double-layer HTSC's in Ref.\ \onlinecite{tokunaga99}, we can estimate $P_{NQR}$ for the reported Knight shift data of Hg1201 and Hg1212.\cite{itoh98} The $P_{NQR}$ of Hg1201 and Hg1212 are also plotted in the same figure as the star and upward triangle, respectively. Except underdoped Hg1201 that show some deviation from the $P_{pl}$ = $P_{NQR}$ line, the $P_{NQR}$ for both samples also show the good correlation with $P_{pl}$. The correspondence between $P_{pl}$ and $P_{NQR}$ suggests that the hole concentration macroscopically determined from TEP is consistent with the hole concentration microscopically determined from NQR. 

\subsection{Electronic phase diagram by RT scale}

\begin{figure*}
\includegraphics[scale=1.5]{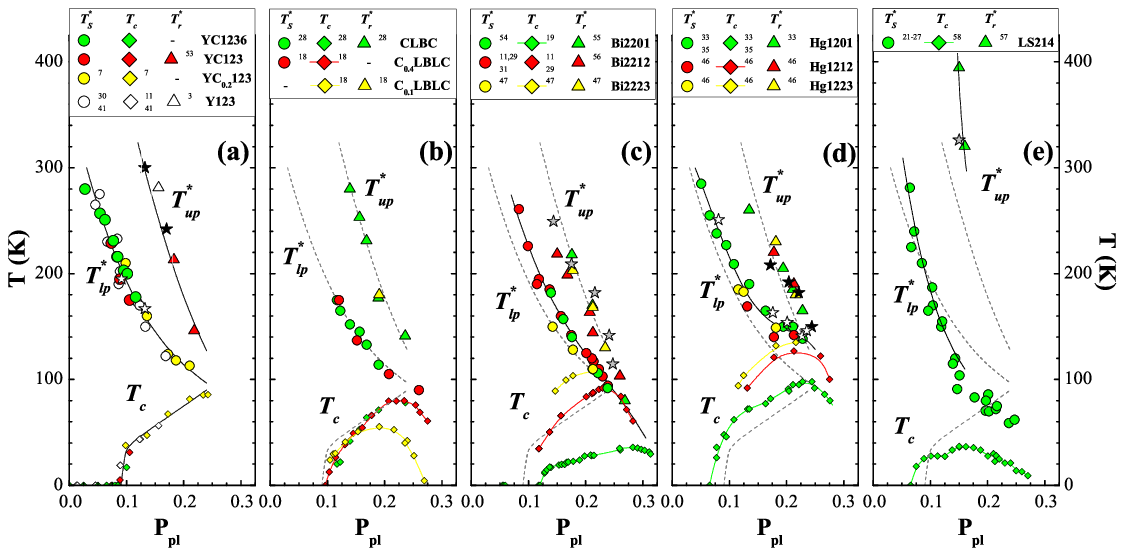}
\caption{\label{fig8}(color online) $T_{up}^*$ and $T_{lp}^*$ vs $P_{pl}$ in (a) Y123 and YC123, (b) CLBC and C$_x$LBLC (c) Bi-based family, (d) Hg-based family and (e) LS214. All solid lines are guide to the eyes for $T\!_{up}$, $T\!_{lp}$ and $T_c$. The dashed lines represent the same curves as the solid lines for Y123 and YC123. In (a) and (d), the open and closed stars show $T_s^*$ and $T_{S2}^*$ for Zn-substituted Y123 and Zn-substituted Hg1201, respectively.\cite{tallon95b,yamamoto02} In (c) and (e), the gray stars show $T_{S2}^*$ for Bi$_2$Sr$_2$Ca$_{1-x}$Pr$_x$Cu$_2$O$_8$ and LS214.\cite{takemura00} }
\includegraphics[scale=1.6]{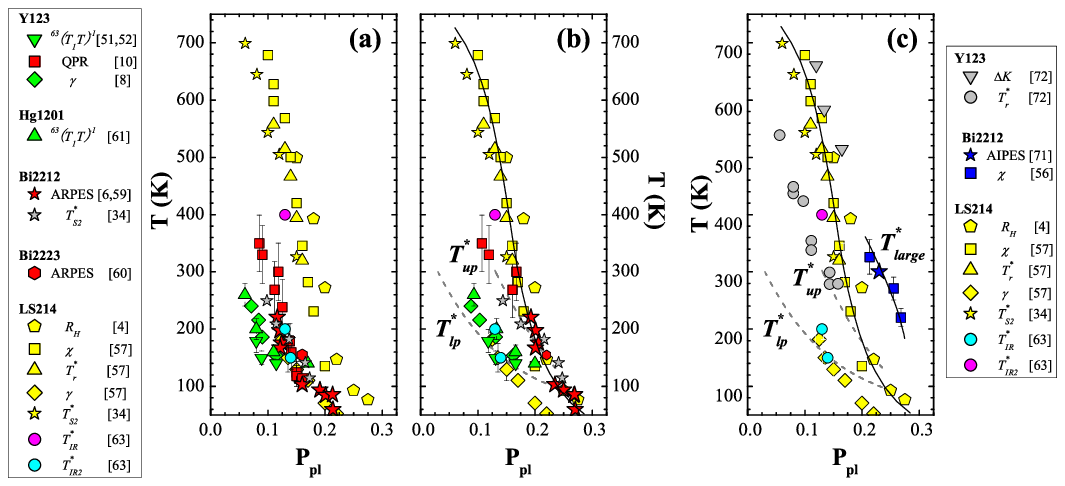}
\caption{\label{fig9}(Color online) Comparison of various characteristic temperatures determined by different probes for different HTSC's as a function of $P_{pl}$ (a) exactly as reported, and (b) and (c) Formula (3a) and (3b). For details see text. The $T_{up}^*$ and $T_{lp}^*$ curves (dashed lines) are same as those in Fig.\ \ref{fig8}. The solid lines are guide to the eyes.}
\end{figure*}

We try to validate our scale by examing the pseudogap behavior in various HTSC's. Since the pseudogap state found in $\rho$, \cite{ito93} ARPES, \cite{ding96} and nuclear magnetic resonance (NMR) experiments \cite{takigawa91} has emerged as an intrinsic property of the electronic states of underdoped HTSC, if plotted on a physically meaningful scale, we expect a universal pseudogap behavior for all HTSC's. 

First of all, the phase diagram for Y$_{1-x}$Ca$_x$Ba$_2$Cu$_3$O$_y$ (YC123) and Y123 is shown in Fig.\ \ref{fig8}(a). $T_S^*$'s for YC1236, YC123,\cite{honma96} Y$_{0.8}$Ca$_{0.2}$Ba$_2$Cu$_3$O$_y$ (YC$_{0.2}$123),\cite{bernhard96} and Y123 \cite{cooper96,cooper91} are found to lie on a common curve which decreases from $\sim$ 300 K at $P_{pl}$ $\approx$ 0.025 to $\sim$ 100 K at $P_{pl}$ $\approx$ 0.24. $T_c$'s for YC1236, YC123,\cite{bernhard96} and Y123 \cite{obertelli92,cooper91} also lie on a common curve which appears at $P_{pl}$ $\approx$ 0.09 and reaches $T_c^{max}$ at $P_{pl}$ $\approx$ 0.25 (curve A in Fig.\ \ref{fig6}). $T_S^*$ seems to be smoothly connected with $T_c$ at slightly overdoped level. $T_S^*$ for both curve A and curve B in Fig.\ \ref{fig6} lay on the same $T_S^*$($P_{pl}$) curve. Another characteristic temperature $T_r^*$, where $\rho$ exhibits a downward deviation from the $T$-linear behavior, for Y123, \cite{ito93} and YC123, \cite{honma96} form another curve above the $T_S^*$ curve. Noted that $T_c$'s for samples with same structure fall on the same curve and therefore, for the same structure, $T_c$ can be used as a secondary measure of $P_{pl}$ if $S^{290}$ is not available. Hereafter, we will call temperatures determined by $T_r^*$ and $T_S^*$ as $upper\ pseudogap\ temperature$ $T_{up}^*$ and $lower\ pseudogap\ temperature$ $T_{lp}^*$, respectively. In Fig.\ \ref{fig8}(b), we plot $T_S^*$, $T_r^*$ and $T_c$ for CLBC\cite{hayashi96} and C$_x$LBLC \cite{knizhnik99} that have crystal structure similar to that of Y123.\cite{goldschmidt93} They show tetragonal symmetry with no ordered chain. $T_S^*$ and $T_r^*$ fall exactly on the $T_{lp}^*$ and $T_{up}^*$ curves found in Y123 and YC123, respectively. Similar $T_{up}^*$ and $T_{lp}^*$ curves can be obtained for Bi-based family of Bi2201, \cite{dumont00,ando99,ando00} Bi2212,\cite{obertelli92,akoshima98,oda97} and Bi$_2$Sr$_2$Ca$_2\!$Cu$_3\!$O$_y$$\!$ (Bi2223) \cite{fujii02} as shown in Fig.\ \ref{fig8}(c) and Hg-based family of Hg1201,\cite{yamamoto00,yamamoto02} Hg1212,\cite{fukuoka97} and Hg1223\cite{fukuoka97} as shown in Fig.\ \ref{fig8}(d), respectively. This clearly suggests that $two$ $pseudogaps$ do not depend on the number of CuO$_2$ layers in the unit cell. However, the superconductivity is enhanced with increasing number of layers, consistent with the empirical rule of $T_c$ for HTSC. The phase diagram for LS214 is shown in Fig.\ \ref{fig8}(e).\cite{nakamura93,nishikawa94,zhou95,kakinuma99,johnston87,xu00,hongjie01,nakano94,radaelli94} It is seen that $both$ $pseudogaps$ follow the universal curves until $P_{pl}$ $\approx$ 0.15 where the optimal $T_c$ is. It is interesting to note that although our universal $P_{pl}$ scale was originally derived from $S^{290}$ of LS214 it does seem that LS214 is an exceptional member of HTSC as commonly believed.

We further demonstrate the possible application and the advantages of using our scale. In Fig.\ \ref{fig9}(a), we plot the spin gap temperature signaled by a decrease in $^{63}\!$Cu nuclear relaxation rate $^{63}\!(T_1T)^{-1}$ with reducing $T$ and the pseudogap temperature suggested by the disappearance of leading-edge gap observed by ARPES as a function of $P_{pl}$ exactly as reported in the literature.\cite{ding96,sato02,takigawa91,goto96,campuzano99,itoh96} There is, except of LS214, no clear distinction among them and they seem to behave as one pseudogap. However if we re-plot the same data using our scale, \cite{there} Fig.\ \ref{fig9}(b), the pseudogap and spin gap clearly belong to the  $T_{up}^*$ and $T_{lp}^*$, respectively.

The characteristic temperatures determined from the time resolved quasiparticle relaxation (QPR) measurement, \cite{kabanov99} $R\!_H$, \cite{hwang94} and magnetic susceptibility $\chi$,\cite{nakano94} that are also included in Fig.\ \ref{fig9}, fall on the $T_{up}^*$ curve. The temperature of a broad peak observed in $T$-dependnce of the electronic specific heat coefficient $\gamma$ falls on the $T_{lp}^*$ curve.\cite{loram93,loram98} Thus, various characteristic temperatures including LS214 belong to either $T_{up}^*$ or $T_{lp}^*$.

Here, we would like to point out that $T_{lp}^*$ and $T_{up}^*$ are not two temperatures, determined by different experimental probes, of a common origin. Rather, they have distinct different physical origins. This can be infer from the suppression of $S$($T$) by Zn substitution below $T_{S2}^*$ ($>$ $T_S^*$). The $T_{S2}^*$'s (filled stars) lie exactly on the $T_{up}^*$ curve as shown in Fig.\ \ref{fig8}(a) and (d), while $T_S^*$ (open stars) for the samples substituted until 1 $\%$ in YBa$_2$(Cu$_{1-z}$Zn$_z$)$_3$O$_y$\cite{tallon95b} and 3 $\%$ in HgBa$_2$(Cu$_{1-z}$Zn$_z$)O$_{4+\delta}$\cite{yamamoto02} fall on the $T_{lp}^*$ curve. Further, the $T_{S2}^*$ determined from the $S$($T$) in Bi$_2$Sr$_2$Ca$_{1-x}$Pr$_x$Cu$_2$O$_8$ and LS214 by another scaling method were plotted into Fig.\ \ref{fig8}(c) and (e) as gray stars, respectively.\cite{takemura00} The same $T_{S2}^*$ for Bi$_2$Sr$_2$Ca$_{1-x}$Pr$_x$Cu$_2$O$_8$ and LS214 also were plotted into Fig.\ \ref{fig9}(a) and (b) as gray and yellow stars, respectively. The $T_{S2}^*$ lie on the upper pseudogap temperature. Therefore both $T_{lp}^*$ and $T_{up}^*$ are sequentially observed by a single TEP measurement. This result strongly suggests that there are two characteristic temperatures. Another experimental result that we can address is infrared response.\cite{startseva99} The frequency-dependent effective scattering rate 1/$\tau$ shows two characteristic temperatures, namely, the temperature $T_{IR}^*$ where the low-frequency 1/$\tau$ starts to be clearly suppressed below 700 cm$^{-1}$ and the temperature $T_{IR2}^*$ where the high-frequency 1/$\tau$ starts to depend on $T$. For LS214, the former corresponds to the $T_{lp}^*$ and the latter corresponds to $T_{up}^*$ as shown in Fig.\ \ref{fig9}(a) and (b).

In the optimally doped La$_{1.85}$Sr$_{0.15}$CuO$_4$ ($P_{pl}$ = 0.15), the ARPES result shows that the pseudogap temperatures were observed beyond $\sim$ 200 K, although the temperature is much higher than $T_c$ = 38 K.\cite{sato99} This is significantly different from previous results.\cite{ding96} However, on our phase diagram as shown in Fig.\ \ref{fig8}(e) and\ \ref{fig9}(b), the unusual pseudogap temperature at $\sim$ 200 K belongs to the upper pseudogap temperature, consistent with that observed by ARPES in Bi2212.

Although there are various discussions for double pseudogaps,\cite{batlogg96,emery97,schmalian98,markiewicz02b,mihailovic99,tanamoto94} our results provide clear experimental evidences and the calibration (temperature versus carrier concentration) curves of two distinct universal intrinsic pseudogaps in HTSC. The physical origins of $upper$ and $lower$ $pseudogaps$ have been attributed to the onset of the electronic inhomogeneity and the superconducting fluctuation, respectively.\cite{emery97} If we adopt this scenario, our results indicate that both electronic inhomogeneity and superconducting fluctuation in the pseudogap regimes are strictly "2D". 

T. Tanamoto et al. noticed the two characteristic temperatures observed in Y123, the temperature at which ($T_1T$)$^{-1}$ versus $T$ takes a maximum, $T_R$, and the onset temperature of the suppression of $^{89}$$\Delta$$K$, $T_S$, and studied both temperatures by the extended $t$-$J$ model.\cite{tanamoto94} On the present scale, the $T_R$ corresponds to the lower pseudogap temperatures.  

According to the recent experimental result of ARPES and the angle-integrated photoemission spectroscopy (AIPES) in Bi2212, there are two types of pseudogap were observed.\cite{takahasi01} One is small pseudogap, the one usually observed by ARPES in Bi2212. The other is large pseudogap observed only by AIPES. While the small pseudogap temperature touches $T_c$ at slightly overdoped level, the large pseudogap temperature is much higher than $T_c$ at same doping level. The large pseudogap by AIPES is plotted in Fig.\ \ref{fig9}(c) as blue stars. In Bi2212, the third pseudogap temperature was reported in $T$-dependence of the uniform magnetic susceptibility.\cite{oda97} Further, in Y123, the onset temperature of the suppression of $\rho$ and $\Delta$$K$ above RT was estimated from the $T$-dependence of $\rho$ and $\Delta$$K$ observed below RT by the scaling.\cite{wuyts96} Their data also are plotted in Fig.\ \ref{fig9}(c). On the present scale, it seems to suggest the possible existence of a third pseudogap. Further experimental and theoretical studies are required to pin down the physical origins of these pseudogaps and their behaviors.

\section{SUMMARY AND CONCLUSIONS}

In summary, we have shown that $S^{290}\!$($P_{pl}$) of double-layer Y$_{1-x}$Ca$_x$Ba$_2$Cu$_3$O$_6$ follows that of single-layer LS214 and can be represented by Formula (3). Although it is not clear exactly how this scale works so well, we argued that $S^{290}$ is dictated by some intrinsic equilibrium properties of the electronic states of doped holes and, therefore, can be used as a common scale to measure $P\!_{pl}$ of layered HTSC cuprates. Indeed, $S^{290}$'s were found, independent of if it is doped with hard or soft dopants, \cite{lorenz02} to be identical in both La$_2\!$CuO$\!_{4+\delta}$ and La$_{2-z}$Sr$_z$CuO$_4$ up to $P_{pl}$ $\approx$ 0.1.\cite{li96} A universal phase diagram of HTSC is constructed by using our proposed scale of $S^{290}$. We conclude that $double$ $pseudogaps$ are intrinsically a single CuO$_2$ layer, therefore 2D in nature, property and $T_c$ depends on the inter-layer coupling for all HTSC's. Most recent experimental results may suggest the existence of a third pseudogap. Our proposed scale points to a unified way to systematically study and compare physical properties of different HTSC and provides further insights of the possible distinct origins of $two$ $pseudogaps$.

\begin{acknowledgments}
One of us (T.H.) would like to thank Dr. K. Yamaya and Dr. K. Kodaira of Hokkaido University, and Dr. S. Yomo of Hokkaido Tokai University for supporting the present research at the initial stage. We are indebted to Dr. Y.S. Song for his technical assistance with TEP measurement. This work was supported by the state of Texas through the Texas Center for Superconductivity at the University of Houston.
\end{acknowledgments}


\begin{thebibliography}{}\label{sec:TeXbooks}
\bibitem{carrington94}A. Carrington and J.R. Cooper, Physica (Amsterdam) {\bf 219C}, 119 (1994): I.R. Fisher and J.R. Cooper, Physica (Amsterdam) {\bf 272C}, 125 (1996). 
\bibitem{yasuoka89}H. Yasuoka, T. Imai and T. Shimizu, in $Strong$ $Correlations$ $and$ $Superconductivity$, edited by H. Fukuyama, S. Maekawa, A.P. Malozemoff, Springer Series in Solid-State Sciences, Vol.89 (Springer-Verlag, Berlin, 1989), p. 254.
\bibitem{ito93}T. Ito, K. Takenaka, S. Uchida, Phys. Rev. Lett. {\bf 70}, 3995 (1993).
\bibitem{hwang94}H.Y. Hwang, B. Batlogg, H. Takagi, H.L. Kao, J. Kwo, R.J. Cava, J.J. Krajewski and W.F. Peck, Jr., Phys. Rev. Lett. {\bf 72}, 2636 (1994).
\bibitem{basov94}D.N. Basov, T. Timusk, B. Dabrowski and J.D. Jorgensen, Phys. Rev. B {\bf 50}, R3511 (1994).
\bibitem{ding96}H. Ding, T. Yokoya, J.C. Campuzano, T. Takahashi, M. Randeria, M.R. Norman, T. Mochiku, K. Kadowaki and J. Giapintzakis, Nature (London) {\bf 382}, 51 (1996)
\bibitem{bernhard96}C.$\!$ Bernhard and J.L.$\!$ Tallon, Phys.$\!$ Rev.$\!$ B$\!$ {\bf 54},$\!$ 10201$\!$ (1996).
\bibitem{loram93}J.W. Loram, K.A. Mirza, J.R. Cooper and W.Y. Liang, Phys. Rev. Lett. {\bf 71}, 1740 (1993).
\bibitem{loram98}J.W. Loram, K.A. Mirza, J.R. Cooper and J.L. Tallon, J. Phys. Chem. Solids {\bf 59}, 2091 (1998).
\bibitem{kabanov99}V.V. Kabanov, J. Demsar, B. Podobnik and D. Mihailovic, Phys. Rev. B {\bf 59}, 1497 (1999).
\bibitem{obertelli92}S.D. Obertelli, J.R. Cooper and J.L. Tallon, Phys. Rev. B {\bf 46}, R14928 (1992).
\bibitem{tallon01}J.L.Tallon and J.W. Loram, Physica (Amsterdam) {\bf 349C}, 53 (2001).
\bibitem{presland91}M.R. Presland, J.L. Tallon, R.G. Buckly, R.S. Liu and N.E. Flower, Physica (Amsterdam) {\bf 176C}, 95 (1991).
\bibitem{tallon90}J.L. Tallon, Physica (Amsterdam) {\bf 168C}, 85 (1990).
\bibitem{tallon95}J.L. Tallon, C. Bernhard, H. Shaked, R.L. Hitterman and J.D. Jorgensen, Phys. Rev. B {\bf 51}, R12911 (1995).
\bibitem{markiewicz02}R.S. Markiewicz and C. Kusko, Phys. Rev. B {\bf 65}, 064520 (2002).
\bibitem{goldschmidt93}D. Goldschmidt, G.M. Reisner, Y. Direktovitch, A. Knizhnik, E. Gartstein, G. Kimmel and Y. Eckstein, Phys. Rev. B {\bf 48}, 532 (1993).
\bibitem{knizhnik99}A.$\!$ Knizhnik$\!$, Y. Direktovich, G.M. Reisner, D. Goldschmidt, C.G. Kuper and Y. Eckstein, Physica$\!$ (Amsterdam)$\!$ {\bf 321C},$\!$ 199$\!$ (1999).
\bibitem{ando00}Y. Ando, Y. Hanaki, S. Ono, T. Murayama, K. Segawa, N. Miyamoto and S. Komiya, Phys. Rev. B {\bf 61}, R14956 (2000).
\bibitem{choi01}E.S. Choi, J.S. Brooks, J.S. Qualls and Y.S. Song, Rev. Sci. Instrum. {\bf 72}, 2392 (2001).
\bibitem{nakamura93}Y. Nakamura and S. Uchida, Phys. Rev. B {\bf 47}, R8369 (1993).
\bibitem{nishikawa94}T. Nishikawa, J. Takeda and M. Sato, J. Phys. Soc. Jpn. {\bf 63}, 1441 (1994).
\bibitem{zhou95}J.-S. Zhou and J.B. Goodenough, Phys. Rev. B {\bf 51}, 3104 (1995).
\bibitem{kakinuma99}N. Kakinuma, Y. Ono and Y. Koike, Phys. Rev. B {\bf 59}, 1491 (1999). 
\bibitem{johnston87}D.C. Johnston, J. P. Stokes, D. P. Goshorn and J. T. Lewandowski, Phys. Rev. B {\bf 36}, R4007 (1987).
\bibitem{xu00}Z.A. Xu, N.P. Ong, T. Kakeshita, H. Eisaki and S. Uchida, Physica (Amsterdam)  {\bf 341-348C} , 1711 (2000).
\bibitem{hongjie01}Y. Hongjie, W. Bin, Z. Lei, L. Ang, M. Zhiqiang, X. Gaojie, Z. Yuheng and W. Ye-ning, Physica (Amsterdam) {\bf 353C}, 221 (2001).
\bibitem{hayashi96}K. Hayashi, K. Matsuura, Y. Okajima, S. Tanda, N. Homma and K. Yamaya, Czech. J. Phys. {\bf 46} (S2), 1171 (1996).
\bibitem{akoshima98}M. Akoshima, T. Noji, Y. Ono and Y. Koike, Phys. Rev. B {\bf 57}, 7491 (1998).
\bibitem{cooper96}J.R. Cooper and J.W. Loram, J. Phys. I France {\bf 6}, 2237 (1996).
\bibitem{mandal96}J.B. Mandal, A.N. Das and B. Gosh, J. Phys. : Conds. Matter {\bf 8}, 3047 (1996).
\bibitem{tallon95b}J.L. Tallon, J.R. Cooper, P.S.I.P.N. de Silva, G.V.M. Williams and J.W. Loram, Phys. Rev. Lett. {\bf 75}, 4114 (1995).
\bibitem{yamamoto00}A. Yamamoto, W.-Z. Hu, and S. Tajima, Phys. Rev. B {\bf 63}, 024504 (2000).
\bibitem{takemura00}T. Takemura, T. Kitajima, T. Sugaya and I. Terasaki, J. Phys. : Conds. Matter {\bf 12}, 6199 (2000).
\bibitem{yamamoto02}A. Yamamoto, K. Minami, W.-Z. Hu, A. Miyakita, M. Izumi and S. Tajima, Phys. Rev. B {\bf 65}, 104505 (2002).
\bibitem{cooper87}J.R. Cooper, B. Alavi, L-W. Zhou, W.P. Beyermann and G. Gr\"{u}ner, Phys. Rev. B {\bf 35}, R8794 (1987).
\bibitem{yu88}R.C. Yu, M.J. Naughton, X. Yan, P.M. Chaikin, F. Holtzberg, R.L. Greene, J. Stuart and P. Davies, Phys. Rev. B {\bf 37}, R7963 (1988).
\bibitem{nagaosa90}N. Nagaosa and P.A. Lee, Phys. Rev. Lett. {\bf 64}, 2450 (1990).
\bibitem{here}Here, we exclude compounds with conducting channels other than CuO$_2$ layer such as chain layer in fully oxygenated YBa$_2$Cu$_3$O$_7$.
\bibitem{honma97}T. Honma, K. Yamaya, N. M\^{o}ri and M. Tanimoto, in $Proceeding$ $of$ $the$ $9th$ $International$ $Symposium$ $on$ $Superconductivity$, $Sapporo$, $1996$, edited by S. Nakajima and M. Murakami (Springer, Tokyo, 1997), p. 253.
\bibitem{cooper91}J.R. Cooper, S.D. Obertelli, A. Carrington and J.W. Loram, Phys. Rev. B {\bf 44}, R12086 (1991).
\bibitem{akoshimab98}M. Akoshima and Y. Koike, J. Phys. Soc. Jpn. {\bf 67}, 3653 (1998).
\bibitem{cooper00}J.R. Cooper, H. Minami, V.W. Wittorff, D. Babi\'{c} and J.W. Loram, Physica (Amsterdam) {\bf 341-348C}, 855 (2000).
\bibitem{wang01}Y. Wang and N.P. Ong, Proc. Nat. Acad. Sci. {\bf 98}, 11091 (2001); cond-mat/0110215v1 (2001).
\bibitem{veal90}B.W. Veal, H. You, A.P. Paulikas, H. Shi, Y. Fang and J.W. Downey, Phys. Rev. B {\bf 42}, 4770 (1990).
\bibitem{fukuoka97}A. Fukuoka, A. Tokiwa-Yamamoto, M. Itoh, R. Usami, S. Adachi and K. Tanabe, Phys. Rev. B {\bf 55}, 6612 (1997).
\bibitem{fujii02}T. Fujii, I. Terasaki, T. Watanabe and A Matsuda, Phys. Rev. B {\bf 66}, 024507 (2002).
\bibitem{tokunaga99}Y. Tokunaga, H. Kotegawa, K. Ishida, G.-q. Zheng, Y. Kitaoka, K. Tokiwa, A. Iyo and H. Ihara, J. Low. Temp. Phys. {\bf117}, 473 (1999)
\bibitem{itoh98}Y. Itoh, T. Machi, S. Adachi, A. Fukuoka, K. Tanabe and H. Yasu,, J. Phys. Soc. Jpn. {\bf 67}, 312 (1998).
\bibitem{tokunaga00}Y. Tokunaga, K. Ishida, Y. Kitaoka, K. Asayama, K. Tokiwa, A. Iyo and H. Ihara, Phys. Rev. B {\bf 61}, 9707 (2000).
\bibitem{takigawa91}M. Takigawa, A.P. Reyes, P.C. Hammel, J.D. Thompson, R.H. Heffner, Z. Fisk and K.C. Ott, Phys. Rev. B {\bf 43}, 247, (1991).
\bibitem{goto96}A. Goto, H. Yasuoka and Y. Ueda, J. Phys. Soc. Jpn. {\bf 65}, 3043 (1996).
\bibitem{honma96}T. Honma, K. Yamaya, N. M\^{o}ri and M. Tanimoto, Solid State Commun. {\bf 98}, 395 (1996).
\bibitem{dumont00}Y. Dumont, C. Ayache and G. Collin, Phys. Rev. B {\bf 62}, 622 (2000).
\bibitem{ando99}Y.$\!$ Ando and T.$\!$ Murayama,$\!$ Phys.$\!$ Rev.$\!$ B$\!$ {\bf 60},$\!$ R6991$\!$ (1999).
\bibitem{oda97}M. Oda, K. Hoya, R. Kubota, C. Manabe, N. Momono, T. Nakano and M. Ido, Physica (Amsterdam) {\bf 281C}, 135 (1997).
\bibitem{nakano94}T. Nakano, M. Oda, C. Manabe, N. Momono, Y. Miura and M. Ido, Phys. Rev. B {\bf 49}, 16000 (1994).
\bibitem{radaelli94}P.G. Radaelli, D.G. Hinks, A.W. Mitchell, B.A. Hunter, J.L. Wagner, B. Dabrowski, K.G. Vandervoort, H.K. Viswanathan and J.D. Jorgensen, Phys. Rev. B {\bf 49}, 4163 (1994).\bibitem{campuzano99} J.C. Campuzano $et\ al.$, Phys. Rev. Lett. {\bf 83}, 3709 (1999).
\bibitem{sato02}T. Sato, H. Matsui, S. Nishina, T. Takahashi, T. Fujii, T. Watanabe and A. Matsuda, Phys. Rev. Lett. {\bf 89}, 067005 (2002).
\bibitem{itoh96}Y. Itoh, T. Machi, A. Fukuoka, K. Tanabe and H. Yasuoka, J. Phys. Soc. Jpn. {\bf 65}, 3751 (1996).
\bibitem{there}Since there is no corresponding $S^{290}$ data except LS214, $P_{pl}$ of all data points plotted in Fig. 9(b) were determined by $"T_c"$, our secondary measure of $P_{pl}$.
\bibitem{startseva99}T. Startseva, T. Timusk, A.V. Puchkov, D.N. Basov, H.A. Mook, M. Okuya, T. Kimura and K. Kishio, Phys. Rev. B {\bf 59}, 7184 (1999).
\bibitem{sato99}T. Sato, T. Yokoya, Y. Naitoh, T. Takahashi, K. Yamada and Y. Endoh, Phys. Rev. Lett. {\bf 83}, 2254 (1999).
\bibitem{batlogg96}B. Batlogg and V.J. Emery, Nature (London) {\bf 382}, 20 (1996).
\bibitem{emery97}V.J. Emery, S.A. Kivelson and O. Zachar, Phys. Rev. B {\bf 56}, 6120 (1997).
\bibitem{schmalian98}J. Schmalian, D. Pines and B. Stojkovi\'{c}, Phys. Rev. Lett. {\bf 80}, 3839 (1998).
\bibitem{markiewicz02b}R.S. Markiewicz, Phys. Rev. Lett. {\bf 89}, 229703 (2002).
\bibitem{mihailovic99} D. Mihailovic, V.V. Kabanov, K. \v{Z}agar and J. Demsar, Phys. Rev. B {\bf 60}, R6995 (1999).
\bibitem{tanamoto94}T. Tanamoto, H. Kohno, H. Fukuyama: J. Phys. Soc. Jpn. {\bf 63}, 2739 (1994).
%\bibitem{andergassen01}S. Andergassen, S. Caprara, C. Di Castro and M. Grilli, Phys. Rev. Lett. {\bf 87}, 056401 (2001).
\bibitem{takahasi01}T. Takahasi, T. Sato, T. Yokoya, T. Kamiyama, Y. Naitoh, T. Mochiku, K. Yamada, Y. Endoh and K. Kadowaki, J. Phys. Chem. Solids {\bf 62}, 41 (2001).
%\bibitem{ishida98}K. Ishida, Y. Yoshida, T. Mito, Y. Tokunaga, Y. Kitaoka and K. Asayama, Phys. Rev. B {\bf 58}, R5960 (1998).
\bibitem{wuyts96}B. Wuyts, V.V. Moshchalkov and Y. Bruynseraede, Phys. Rev. B {\bf 53}, 9418 (1996).
%\bibitem{williams96}G.V.M. Williams, J.L. Tallon, R. Michalak and R. Dupree, Phys. Rev. B {\bf 54}, R6909 (1996).
\bibitem{lorenz02}B. Lorenz, Z.G. Li, T. Honma and P.H. Hor, Phys. Rev. B {\bf 65}, 144522 (2002).
\bibitem{li96}Z.G. Li, H.H. Feng, Z.Y. Yang, A. Hamed, S.T. Ting, P.H. Hor, S. Bhavaraju, J.F. DiCarlo and A.J. Jacobson, Phys. Rev. Lett. {\bf 77}, 5413 (1996).





\end{thebibliography}
\end{document}